\documentclass[12pt]{article}
\usepackage{amsfonts}
\newcommand{\text}{\rm}

\begin{document}

\title{\textbf{Seiberg-Witten map for the 4D noncommutative BF theory}}
\author{L. C. Q. Vilar, O.S. Ventura$^{a,b}$, R. L. P. G. Amaral$^{c}$, \\ V. E. R. Lemes$^{d}$  and L. O. Buffon$^{b,e}$ 
\footnote{mbschlee@terra.com.br, ozemar@cefetes.br, rubens@if.uff.br, vitor@dft.if.uerj.br, lobuffon@terra.com.br}\\
{ \small \em $^a$Coordenadoria de F\'{\i}sica, Centro Federal de Educa\c {c}\~{a}o Tecnol\'{o}gica \  do Esp\'{\i}rito Santo},  \\
\small\em $ $Avenida Vit\'{o}ria 1729 - Jucutuquara, Vit\'{o}ria - ES, 29040 - 333, Brazil\\
\small\em $^b$Centro Universit\'{a}rio de Vila Velha, Rua comiss\'{a}rio Jos\'{e} \\
\small\em Dantas de Mello 15 - Boa Vista, Vila Velha - ES, 29102 - 770, Brazil\\
\small\em $^c$Instituto de F\'\i sica, Universidade Federal Fluminense, 24210 - 340, Niter\'oi - RJ, Brasil,\\
\small\em $^d$Instituto de F\'\i sica, Universidade do Estado do Rio de Janeiro,\\ 
\small\em Rua S\~{a}o Francisco Xavier 524, Maracan\~{a}, Rio de Janeiro - RJ, 20550-013, Brazil \\
\small\em $^e$Escola Superior de Ci\^{e}ncias da Santa Casa de Miseric\'{o}rdia de Vit\'{o}ria,\\ \small\em Av. Nossa Senhora da Penha 2190, Santa Luiza, Vit\'{o}ria-ES, 29045-402, Brazil. }
\bigskip
\maketitle

\vspace{-1cm}
\begin{abstract}
 We describe the Seiberg-Witten map taking the 4D noncommutative BF theory (NCBF) into its pure commutative version.The existence of this map is in agreement with the hypothesis that such maps are available for any noncommutative theory with Schwarz type topological sectors, and represents a strong indicative for the renormalizability of these theories in general.

\end{abstract}
\setcounter{page}{0}\thispagestyle{empty}

\vfill\newpage\ \makeatother

\section{Introduction}

Distinct gauge choices in the open strings lead to the realization of both
ordinary Yang-Mills field theories as well as noncommutative field theories.
It was the perception of this fact that made Seiberg and Witten propose the
so called Seiberg-Witten map (SW map) \cite{sw}. This mapping establishes an
expression of noncommutative field variables in terms of ordinary
(commutative) fields, in such a way that the noncommutative gauge
transformed fields are mapped into commutative fields gauge transformed in
the ordinary sense.

Let us briefly review this idea. First we introduce the Moyal product
between two functions defined on the noncommutative space \cite{moyal}:

\begin{eqnarray}
f*g &=&\exp \left( \frac i2\theta ^{ij}\frac \partial {\partial x^i}\frac
\partial {\partial y^j}\right) f\left( x\right) g\left( y\right)_{y\rightarrow x}  \nonumber  \label{bffgh} \\
&=&fg+\frac i2\theta ^{ij}\partial _if\partial _jg+O\left( \theta ^2\right) ,
\label{moyal}
\end{eqnarray}

\noindent where the real c-number parameter $\theta^{ij}$ is originated in
the noncommutativity of space-time coordinates 
\begin{equation}
\left[ x^i,x^j\right] =i\theta ^{ij}.  \label{theta}
\end{equation}

\noindent The vanishing of $\theta$ turns the noncommutative theory into the
commutative one. Then, after defining the Moyal bracket, 
\begin{equation}
\left[ f\stackrel{*}{,}g\right] =f*g-g*f,  \label{moyal com}
\end{equation}
the noncommutative gauge transformation is constructed as

\begin{equation}
\stackrel{\wedge }{\delta }_{\stackrel{\wedge }{\lambda }}\stackrel{\wedge }{A}
\,=\partial _i\stackrel{\wedge }{\lambda }+\,i\left[ \stackrel{\wedge }{
\lambda }\, \stackrel{*}{,}\, \stackrel{\wedge }{A_i}\right] =\, \stackrel{
\wedge }{D}_i\stackrel{\wedge }{\lambda }.  \label{transf}
\end{equation}

\noindent Here the hat symbol identifies fields and operators defined on the
noncommutative space-time while $\stackrel{\wedge }{D}_{i}$ represents the
Moyal covariant derivative. At this point it should be observed that,
although similar in form to a nonabelian gauge transformation, there is a
nonvanishing contribution coming from the Moyal bracket in (\ref{transf})
even for an abelian gauge field. In the same way, the noncommutative gauge
curvature

\begin{equation}
\stackrel{\wedge }{F}_{ij}=\partial _{i}\stackrel{\wedge }{A_{j}}-\partial _{j}%
\stackrel{\wedge }{A_{i}}-i\left[ \stackrel{\wedge }{A_{i}}\,\stackrel{\ast }{,}%
\,\stackrel{\wedge }{A_{j}}\right]  \label{fij}
\end{equation}
presents a nonvanishing commutator contribution even for an abelian field.

In the abelian case the above expressions, up to first order in $\theta$,
turn out to be

\begin{equation}
\stackrel{\wedge }{\delta }_{\stackrel{\wedge }{\lambda }}\stackrel{\wedge }{A}%
_{i}\,=\partial _{i}\stackrel{\wedge }{\lambda }-\theta ^{kl}\partial _{k}%
\stackrel{\wedge }{\lambda }\partial _{l}\stackrel{\wedge }{A}_{i}+O\left(
\theta ^{2}\right) ,  \label{delta a ex}
\end{equation}

\begin{equation}
\stackrel{\wedge }{F}_{ij}=\partial _{i}\stackrel{\wedge }{A_{j}}-\partial _{j}%
\stackrel{\wedge }{A_{i}}+\theta ^{kl}\partial _{k}\stackrel{\wedge }{A}%
_{i}\partial _{l}\stackrel{\wedge }{A}_{j}+O\left( \theta ^{2}\right) .
\label{fij ex}
\end{equation}

Now, the sense of the SW map is that noncommutative gauge equivalent fields
should be mapped into ordinary gauge equivalent fields. A particular
solution for this problem in the $U(1)$ case, again up to first order in $%
\theta$, is given by \cite{sw}

\begin{equation}
\stackrel{\wedge }{A}_{i}\left( A\right) =A_{i}-\theta ^{kl}\left(
A_{k}\partial _{l}A_{i}-\frac{1}{2}A_{k}\partial _{i}A_{l}\right) +O\left(
\theta ^{2}\right) \mathrm{{\ }{,}}  \label{asw}
\end{equation}

\begin{equation}
\stackrel{\wedge }{\lambda }\left( \lambda ,A\right) =\lambda +\frac{1}{2}%
\theta ^{kl}\left( \partial _{k}\lambda \right) A_{l}+O\left( \theta
^{2}\right) \,,  \label{lsw}
\end{equation}%
and a cohomological characterization of the general solution was
accomplished in \cite{nosso}. As a by-product of this study, a conjecture on
the existence of SW maps taking noncommutative theories in the presence of
topological terms into renormalizable commutative field theories was
proposed. The case of $3D$ theories was analysed \ in detail. It was shown
how a map of the $3D$ noncommutative Maxwell theory in the presence of a
noncommutative Chern-Simons term ($NCCSM$) would lead to the usual
commutative theory \cite{nosso}, whereas $\theta -$dependent nonrenormalizable
interactions inevitably accompany the SW mapped $NCM$ theory in the absence
of a topological sector \cite{Liu}.

In fact, the lack of renormalizability of noncommutative gauge theories already appeared at the noncommutative space, when it was noticed that infrared and ultraviolet divergencies were deeply entangled \cite{matusis,minwalla}. So it could not be a surprise at all to find that the SW mapping of such theories led to commutative theories presenting those $\theta$ dependent nonrenormalizable couplings. Since then, some evolution has been made in this topic. Grosse and Wulkenhaar proposed to couple oscilator like terms to the original noncommutative theories in order to avoid the IR/UV mixing \cite{Grosse}. Slavnov also proposed the coupling of a new term to render the gauge action renormalizable \cite{Slavnov}. Recently, it was shown that this new term was effectively a BF like coupling in the noncommutative space, and that the renormalizability so acchieved should be expected due to the strenght of the supersymmetry present in topological theories \cite{Blaschke}. This is in agreement with the idea that a SW map can always join noncommutative theories with topological sectors together with the renormalizable commutative theories. 
But, about this point, some evidences pointing towards different conclusions can be found in the literature. Blasi and Maggiori argued for quantum instabilities in $2D$ $NCBF$ theory and conjectured that this would be a general feature of noncommutative theories in higher dimensions \cite{Blasi1}, although in $3D$ they agreed that the ``instabilities'' of $NCCS$ would be harmless as they were always restricted to the trivial BRST sector \cite{Blasi2}, thus supporting again the hypotesis that the renormalizability is connected to the existence of a SW mapped renormalizable commutative version of the original noncommutative theory, as it happens with $NCCS$ \cite{gs}. These theories, as $NCCS$, even having non power-counting couplings in the noncommutative space, would be renormalizable in a broader sense \cite{gomis,Blasi3}. 

So, it is an open question wether in $4D$ the $NCBF$ theory is renormalizable and if there is a SW map taking this theory into a renormalizable commutative one. In view of the breakdown of the renormalizability of $4D$ $NCYM$ theories (and also remembering that commutative $YM$ theory has already been consistently proposed to be seen as a deformation of a topological BF theory \cite{fucito}) answering these questions become relevant to the fate of noncommutative $4D$ theories. Here we wants to give a definitive answer to the second question, showing the existence of the SW map for $NCBF$ theory into commutative pure $BF$ theory without any nonrenormalizable coupling, in much the same way as it happens in the $3D$ $NCCS$ case \cite{gs,nosso}. 

In the next section we will show that the noncommutative space allows the existence of new symmetries not reducible to commutative ones. These symmetries are necessary to construct the SW map of the $NCBF$ theory, and we give the example of this map in the first order in $\theta$. In section 3 we generalize the reasoning of the first order by means of a $BRST$ technique, proving the existence of the SW map of the $NCBF$ theory to all orders. In the conclusion, we show that this map takes the $NCBF$ action into the pure commutative $BF$ action, without the presence of any nonrenormalizable couplings. This supports the conjecture 
on the renormalizability of the SW mapped theories coming from noncommutative theories with topological sectors in $4D$.

\section{The first order SW map}

Let us begin by writing the action
for the $NCBF$ theory%
\begin{equation}
\mathcal{S}_{\widehat{BF}}=\frac{1}{4}\int d^{4}x\varepsilon ^{\mu \nu
\varrho \sigma }\stackrel{\wedge }{B}_{\mu \nu }\ast \stackrel{\wedge }{F}%
_{\varrho \sigma }\mathrm{.}  \label{ncbf}
\end{equation}%
The map that we will search, $\stackrel{\wedge }{A}=\stackrel{\wedge }{A}(A,B)$
and $\stackrel{\wedge }{B}=\stackrel{\wedge }{B}(A,B)$, should now preserve the
gauge symmetry 
\begin{equation}
\stackrel{\wedge }{\delta }_{\stackrel{\wedge }{\lambda }}\stackrel{\wedge }{A}%
_{\mu }\left( A,B\right) =\delta _{\lambda }\left( \stackrel{\wedge }{A}_{\mu
}\left( A,B\right) \right) ,\ \ \ \ \ \ \ \stackrel{\wedge }{\delta }_{%
\stackrel{\wedge }{\lambda }}\stackrel{\wedge }{B}_{\mu \nu }\left( A,B\right)
=\delta _{\lambda }\left( \stackrel{\wedge }{B}_{\mu \nu }\left( A,B\right)
\right) ,  \label{swmap}
\end{equation}%
as well as the topological symmetry typical of $BF$ theories%
\begin{equation}
\stackrel{\wedge }{\delta }_{\stackrel{\wedge }{\psi }}\stackrel{\wedge }{A}%
_{\mu }\left( A,B\right) =\delta _{\psi }\left( \stackrel{\wedge }{A}_{\mu
}\left( A,B\right) \right) ,\ \ \ \ \ \ \ \stackrel{\wedge }{\delta }_{%
\stackrel{\wedge }{\psi }}\stackrel{\wedge }{B}_{\mu \nu }\left( A,B\right)
=\delta _{\psi }\left( \stackrel{\wedge }{B}_{\mu \nu }\left( A,B\right)
\right) ,  \label{swtop}
\end{equation}%
for both fields simultaneously, where, in the abelian case, 
\begin{equation}
\stackrel{\wedge }{\delta }_{\stackrel{\wedge }{\lambda }}\stackrel{\wedge }{A}%
_{\mu }=\stackrel{\wedge }{D}_{\mu }\stackrel{\wedge }{\lambda }\mathrm{{\ },\
\ \ \ \ \stackrel{\wedge }{\delta }_{\stackrel{\wedge }{\lambda }}\stackrel{%
\wedge }{B}_{\mu \nu }=i\left[ \stackrel{\wedge }{\lambda }\stackrel{\ast }{,}%
\stackrel{\wedge }{B}_{\mu \nu }\right] ,}  \label{ncg}
\end{equation}%
\begin{equation}
{\delta }_{\lambda }{A}_{\mu }={\partial }_{\mu }{\lambda }{}\mathrm{{\ },\
\ \ \ \ \ \ \ \ \ \ \ \delta _{\lambda }B_{\mu \nu }=0,}  \label{g}
\end{equation}%
\noindent and

\begin{equation}
\stackrel{\wedge }{\delta }_{\stackrel{\wedge }{\psi }}\stackrel{\wedge }{A}%
_{\mu }=0\ ,\ \ \ \ \ \ \stackrel{\wedge }{\delta }_{\stackrel{\wedge }{\psi }}%
\stackrel{\wedge }{B}_{\mu \nu }=\stackrel{\wedge }{D}_{\mu }\stackrel{\wedge }{%
\psi }_{\nu }-\stackrel{\wedge }{D}_{\nu }\stackrel{\wedge }{\psi }_{\mu }%
\mathrm{{\ },}  \label{nctop}
\end{equation}%
\begin{equation}
{\delta }_{\psi }\mathrm{\ }A_{\mu }=0\ ,\ \ \ \ \ \ \ \ \delta _{\psi
}B_{\mu \nu }=\partial _{\mu }\psi _{\nu }-\partial _{\nu }\psi _{\mu }\ .
\label{top}
\end{equation}

The problem that emerges when we try to construct the Seiberg-Witten map of
the noncommutative BF model is the impossibility of the simultaneous
implementation of the conditions shown in $(\ref{swmap})$ and $(\ref{swtop})$%
. The solution of the SW condition for the $\stackrel{\wedge }{A}_{\mu }$ 
field is actually straightforward since $\stackrel{\wedge }{\delta }_{\stackrel%
{\wedge }{\psi }}\stackrel{\wedge }{A}_{\mu }=0$. The SW mapping for this
sector reduces to the SW problem of purely vectorial field theories. The
cohomological treatment for this specific case was presented by us in \cite%
{nosso}. Then, the main difficulty is in the search for solutions of the $%
\stackrel{\wedge }{B}_{\mu \nu }$ field mapping , and, of course, this
difficulty is related to the topological symmetry (Indeed a treatment that
does not take into account the topological symmetry was presented in \cite%
{Benaoum}. There, the addition to $(\ref{ncbf})$ of a quadratic term of the
type $\stackrel{\wedge }{B}_{\mu \nu }\stackrel{\wedge }{B}{}^{\mu \nu }$
explicitly broke the topological symmetry, turning it possible the
implementation of the usual SW mapping).

So, we start our search by noticing that the condition $(\ref{swtop})$,
although natural, is not the most general one. The presence of a
noncommutative structure allows an infinite set of new symmetries that can
be used to extend the $\stackrel{\wedge }{\delta }_{\stackrel{\wedge }{\psi }}%
\stackrel{\wedge }{B}$ symmetry and then deform $(\ref{swtop})$. \ The
extended symmetry remains an invariance of the noncommutative action $(\ref%
{ncbf})$.

For example, in the first explicit order in $\theta $, the following
operation 
\begin{equation}
\stackrel{\wedge }{\delta }_{\stackrel{\wedge }{^{T}}}\stackrel{\wedge }{B}%
_{\mu \nu }=-\frac{1}{2}\theta ^{\alpha \beta }\left\{ \stackrel{\wedge }{D}%
_{\alpha }\stackrel{\wedge }{F}_{\mu \nu }\stackrel{\ast }{,}\stackrel{\wedge }{%
\Psi }_{\beta }\right\} {+}\frac{1}{4}\theta ^{\alpha \beta }\left\{ \stackrel%
{\wedge }{F}_{[\mu \alpha }\stackrel{\ast }{,}\stackrel{\wedge }{D}_{\beta }%
\stackrel{\wedge }{\Psi }_{\nu ]}-\stackrel{\wedge }{D}_{\nu ]}\stackrel{\wedge 
}{\Psi }_{\beta }\right\} ,\ \ \ \ \ \ \ \stackrel{\wedge }{\delta }_{\stackrel%
{\wedge }{^{T}}}\stackrel{\wedge }{A}_{\mu }=0\mathrm{,}  \label{newtop}
\end{equation}%
generates a symmetry of $(\ref{ncbf})$, 
\begin{equation}
\stackrel{\wedge }{\delta }_{\stackrel{\wedge }{^{T}}}\mathcal{S}_{\widehat{BF}%
}=0\mathrm{{\ }.}
\end{equation}%
It is important to remark that this new symmetry $(\ref{newtop})$ was picked
up among the set of possible symmetries with one $\theta $ because it has a
parameter $\stackrel{\wedge }{\Psi }$ which fits the same dimensionality of
the parameter $\stackrel{\wedge }{\psi }$ of the topological symmetry $(\ref%
{nctop})$. As it will be seen, this is the \textquotedblleft
viable\textquotedblright\ deformation of $(\ref{swtop})$ that we were
looking for. Then, by identifying both parameters, we propose that, at least
in first order in $\theta $, the SW condition should relate this extended
noncommutative topological symmetry%
\begin{equation}
\stackrel{\wedge }{\delta }_{\stackrel{\wedge }{\Psi }}\stackrel{\wedge }{B}%
_{\mu \nu }=\stackrel{\wedge }{\delta }_{\stackrel{\wedge }{\psi }}\stackrel{%
\wedge }{B}_{\mu \nu }+\stackrel{\wedge }{\delta }_{\stackrel{\wedge }{^{T}}}%
\stackrel{\wedge }{B}_{\mu \nu }=\stackrel{\wedge }{D}_{[\mu }\stackrel{\wedge }%
{\Psi }_{\nu ]}-\frac{1}{2}\theta ^{\alpha \beta }\left\{ \stackrel{\wedge }{D%
}_{\alpha }\stackrel{\wedge }{F}_{\mu \nu }\stackrel{\ast }{,}\stackrel{\wedge }%
{\Psi }_{\beta }\right\} {+}\frac{1}{4}\theta ^{\alpha \beta }\left\{ 
\stackrel{\wedge }{F}_{[\mu \alpha }\stackrel{\ast }{,}\stackrel{\wedge }{D}%
_{\beta }\stackrel{\wedge }{\Psi }_{\nu ]}-\stackrel{\wedge }{D}_{\nu ]}%
\stackrel{\wedge }{\Psi }_{\beta }\right\} \mathrm{,}  \label{newnctop}
\end{equation}%
and the standard commutative topological symmetry of $(\ref{top})$, $i.e.$%
\begin{equation}
\stackrel{\wedge }{\delta }_{\stackrel{\wedge }{\Psi }}\stackrel{\wedge }{B}%
_{\mu \nu }\left( A,B\right) =\delta _{\psi }\left( \stackrel{\wedge }{B}%
_{\mu \nu }\left( A,B\right) \right) \mathrm{{\ }.}  \label{newswtop}
\end{equation}%
Now, this condition has a solution at first order in $\theta $ that can be
presented as

\begin{equation}
\stackrel{\wedge }{B}_{\mu \nu }^{1}=-\theta ^{\alpha \beta }A_{\alpha
}\partial _{\beta }B_{\mu \nu }-\frac{1}{2}\theta^{\alpha \beta}F_{[\mu
\alpha }B_{\beta \nu ]}  \label{b1order}
\end{equation}

\begin{equation}
\stackrel{\wedge }{\Psi }_{\mu }^{1}=-\theta ^{\alpha \beta }A_{\alpha
}\partial _{\beta }\psi _{\mu }+\theta ^{\alpha \beta }F_{\alpha \mu }\psi
_{\beta }  \label{psi1order}
\end{equation}

But it is not difficult to see that, at second order in $\theta $, the new
condition $(\ref{newswtop})$ has again an obstruction. Using the same
reasoning, one can search for another symmetry of the $\widehat{BF}$ theory
with a parameter with the same dimensions as $\stackrel{\wedge }{\psi }$, now
at the (explicit) second order in $\theta $; then one can extend the
symmetry $(\ref{newnctop})$ once more, and finally construct a new condition
substituting $(\ref{newswtop})$, which will be solvable at second order.
This can be achieved, but it is not surprising at all to understand that in
the end the obstruction in the solution of the SW map for the topological
symmetry will move to the next order, and so on.

In this scenario, the sensible question is if it is possible to assure the
existence of an extended noncommutative topological symmetry at all orders,
from which one could extract the definite SW map of the noncommutative $%
\widehat{BF}$ theory, and then find its form in the commutative space.

In the following we will prove the existence of such map.

\section{SW map to all orders}

The starting point is the BRST transformations of the noncommutative fields
and ghosts. Analogously to the nonabelian BF case \cite{livro}, they are
given by
\begin{eqnarray}
\stackrel{\wedge }{s}\stackrel{\wedge }{A}_{\mu } &=&\stackrel{\wedge }{D}_{\mu
}\stackrel{\wedge }{c}\mathrm{{\ },}  \label{setbrs} \\
\stackrel{\wedge }{s}\stackrel{\wedge }{c} &=&i\stackrel{\wedge }{c}\ast 
\stackrel{\wedge }{c}\mathrm{{\ },}  \nonumber \\
\stackrel{\wedge }{s}\stackrel{\wedge }{B}_{\mu \nu } &=&\stackrel{\wedge }{D}%
_{[\mu }\stackrel{\wedge }{\psi }_{\nu ]}+i\left[ \stackrel{\wedge }{c}\stackrel%
{\ast }{,}\stackrel{\wedge }{B}_{\mu \nu }\right] \mathrm{{\ },}  \nonumber \\
\stackrel{\wedge }{s}\stackrel{\wedge }{\psi }_{\mu } &=&\stackrel{\wedge }{D}%
_{\mu }\stackrel{\wedge }{\rho }+i\left[ \stackrel{\wedge }{c}\stackrel{\ast }{,%
}\stackrel{\wedge }{\psi }_{\mu }\right] \mathrm{{\ },}  \nonumber \\
\stackrel{\wedge }{s}\stackrel{\wedge }{\rho } &=&i\left[ \stackrel{\wedge }{c}%
\stackrel{\ast }{,}\stackrel{\wedge }{\rho }\right] \mathrm{{\ }.}  \nonumber
\end{eqnarray}

\noindent Let us mention that the ghost $\widehat{\rho }$ is necessary due
to the zero modes in the transformation of $\stackrel{\wedge }{\psi }_{\mu }$.

In this BRST context, the SW map is the solution of the condition relating
the noncommutative BRST transformations and those of the commutative theory.
For example, the conditions for $\stackrel{\wedge }{A}_{\mu }$ and $\stackrel{%
\wedge }{c}$,%
\begin{equation}
\stackrel{\wedge }{s}\stackrel{\wedge }{A}_{\mu }\left( A\right) =s\left( 
\stackrel{\wedge }{A}_{\mu }\left( A\right) \right) \mathrm{{\ },}
\label{swa}
\end{equation}
\begin{equation}
\stackrel{\wedge }{s}\stackrel{\wedge }{c}\left( A,c\right) =s\left( \stackrel{%
\wedge }{c}\left( A,c\right) \right) \mathrm{{\ },}  \label{swc}
\end{equation}%
have the same structure of the pure noncommutative Maxwell case. Then, we
can expect the same solution as we anticipated in $(\ref{asw})$ and $(\ref%
{lsw})$. As we are searching for an argument valid for all orders, let us
derive the relations expressing the full dependence of the mapped fields on $%
\theta $. The commutative operator $s$ obviously does not depend on $\theta $%
, and we can establish the commutation between $s$ and the operator of the
variation under $\theta $,
\begin{equation}
\left[ s,\delta _{\theta }\right] =0\mathrm{{\ }.}  \label{sd}
\end{equation}%
Now, from this relation, the SW conditions $(\ref{swa},\ref{swc})$, and the
BRST transformations $(\ref{setbrs})$, we see that the SW mapped fields $%
\stackrel{\wedge }{c}$ and $\stackrel{\wedge }{A}_{\mu }$ should satisfy
\begin{equation}
\stackrel{\wedge }{s}\left( \delta _{\theta }\stackrel{\wedge }{c}\right)
=i\left\{ \delta _{\theta }\stackrel{\wedge }{c}\stackrel{\ast }{,}\stackrel{%
\wedge }{c}\right\} -\frac{1}{4}\delta \theta ^{\alpha \beta }\left[
\partial _{\alpha }\stackrel{\wedge }{c}\stackrel{\ast }{,}\partial _{\beta }%
\stackrel{\wedge }{c}\right] \mathrm{{\ },}  \label{sdc}
\end{equation}%
\begin{equation}
\stackrel{\wedge }{s}\left( \delta _{\theta }\stackrel{\wedge }{A}_{\mu
}\right) =\stackrel{\wedge }{D}_{\mu }\left( \delta _{\theta }\stackrel{\wedge 
}{c}\right) +i\left[ \delta _{\theta }\stackrel{\wedge }{c}\stackrel{\ast }{,}%
\stackrel{\wedge }{A}_{\mu }\right] +i\left[ \stackrel{\wedge }{c}\stackrel{%
\ast }{,}\delta _{\theta }\stackrel{\wedge }{A}_{\mu }\right] -\frac{1}{2}%
\delta \theta ^{\alpha \beta }\left[ \partial _{\alpha }\stackrel{\wedge }{c}%
\stackrel{\ast }{,}\partial _{\beta }\stackrel{\wedge }{A}_{\mu }\right] 
\mathrm{{\ }.}  \label{sda}
\end{equation}%
A particular solution to this system is given by

\begin{equation}
\delta _{\theta }\stackrel{\wedge }{c}=\frac{1}{4}\delta \theta ^{\alpha
\beta }\{\partial _{\alpha }\stackrel{\wedge }{c}\stackrel{\ast }{,}\stackrel{%
\wedge }{A}_{\beta }\}\mathrm{{\ },}  \label{deltac}
\end{equation}%
\begin{equation}
\delta _{\theta }\stackrel{\wedge }{A}_{\mu }=-\frac{1}{4}\delta \theta
^{\alpha \beta }\{\stackrel{\wedge }{A}_{\mu }\stackrel{\ast }{,}\partial
_{\beta }\stackrel{\wedge }{A}_{\mu }+\stackrel{\wedge }{F}_{\beta \mu }\}%
\mathrm{{\ }.}  \label{deltaa}
\end{equation}%
These solutions were written for the first time in \cite{sw}. As stressed
above, they just represent particular solutions to the system $(\ref{sdc},%
\ref{sda})$. In fact, once the nilpotency of the BRST operator $\stackrel{%
\wedge }{s}$ is assured, the complete solution of the problem requires a
characterization of the cohomological classes of $\stackrel{\wedge }{s}$ with
the convenient quantum numbers. A general analysis (which is not of our
concern here) should follow the lines of the work in \cite{nosso}. For the
moment, we would like to call attention to the fact that the solutions $(\ref%
{deltac})$ and $(\ref{deltaa})$ are defined modulo covariant elements under $%
\stackrel{\wedge }{s},$ $i.e.,$ objects that transform as

\begin{equation}
\stackrel{\wedge }{s}\stackrel{\wedge }{X}=i[\stackrel{\wedge }{c}\stackrel{\ast 
}{,}\stackrel{\wedge }{X}]\mathrm{{\ }.}  \label{cov}
\end{equation}%
Notice also that equations $(\ref{asw})$ and $(\ref{lsw})$ can be seen as
first order solutions of $(\ref{deltac})$ and $(\ref{deltaa})$.

This procedure can now be followed for the other fields and ghosts of the $%
\widehat{BF}$ theory appearing in $(\ref{setbrs})$. The solutions for the $%
\theta $ variations of the ghosts $\stackrel{\wedge }{\rho }$ and $\stackrel{%
\wedge }{\psi }_{\mu }$ are obtained in an analogous and straightforward way
and are given by

\begin{equation}
\delta _{\theta }\stackrel{\wedge }{\rho }=-\frac{1}{4}\delta \theta ^{\alpha
\beta }\left\{ \stackrel{\wedge }{A}_{\alpha }\stackrel{\ast }{,}(\partial
_{\beta }+\stackrel{\wedge }{D}_{\beta })\stackrel{\wedge }{\rho }\right\} 
\mathrm{{\ },}  \label{deltarho}
\end{equation}%
\begin{equation}
\delta _{\theta }\stackrel{\wedge }{\psi }_{\mu }=-\frac{1}{4}\delta \theta
^{\alpha \beta }\left\{ \stackrel{\wedge }{A}_{\alpha }\stackrel{\ast }{,}%
(\partial _{\beta }+\stackrel{\wedge }{D}_{\beta })\stackrel{\wedge }{\psi }%
_{\mu }\right\} +\frac{1}{2}\delta \theta ^{\alpha \beta }\left\{ \stackrel{%
\wedge }{F}_{\alpha \mu }\stackrel{\ast }{,}\stackrel{\wedge }{\psi }_{\beta
}\right\} \mathrm{{\ }.}  \label{deltapsi}
\end{equation}

It is worthwhile to observe now that the above procedure which follows from $(\ref{sd})$ is a systemathical method for the study of the existence of the SW map for a general set of transformations, allowing then the obtention of the explicit SW maps of the fields to any order in $\theta^{\mu\nu}$.

The situation of the $\stackrel{\wedge }{B}_{\mu \nu }$ field is more
involved, and we will analyse it in detail. The problem that we will
describe is in fact the root of all the difficulties that we have been
meeting since the beginning of our work. As we just mentioned, the
nilpotency of the BRST operator $\stackrel{\wedge }{s}$ is the fundamental
pillar of all this construction. This can be understood by remembering that
the SW condition on $\stackrel{\wedge }{B}_{\mu \nu }$ leads to%
\begin{equation}
\stackrel{\wedge }{s}^{2}\stackrel{\wedge }{B}_{\mu \nu }\left( A,B\right)
=s^{2}\left( \stackrel{\wedge }{B}_{\mu \nu }\left( A,B\right) \right) 
\mathrm{{\ },}  \label{s2swmap}
\end{equation}%
and as the BRST operator $s$ is nilpotent on all fields of the abelian $BF$
theory, the existence of the SW map becomes conditioned to the nilpotency of 
$\stackrel{\wedge }{s}$ as well. But the resemblance of the BRST
transformations $(\ref{setbrs})$ on the set of transformations of the
commutative nonabelian $BF$ case brings in a well known problem of the
later: the lack of nilpotency of the BRST operator for the nonabelian $BF$
system. This only comes into play now because it  happens precisely on the $%
\stackrel{\wedge }{B}_{\mu \nu }$ field,%
\begin{equation}
\stackrel{\wedge }{s}^{2}\stackrel{\wedge }{B}_{\mu \nu }=-i\left[ \stackrel{%
\wedge }{F}_{\mu \nu }\stackrel{\ast }{,}\stackrel{\wedge }{\rho }\right] 
\mathrm{{\ }.}  \label{nilbreak}
\end{equation}%
In the quantum treatment of the commutative nonabelian case this is overcome
using the Batalin-Vilkovisky procedure by introducing a term of higher order
in the antifields in the fully quantized action \cite{livro,bv}.Or, in the BRST language, noticing that we are dealing with a topological field theory, we have at our disposal a complete ladder structure \cite{ladder} which makes it imediate the correction of the BRST transformation of $\stackrel{\wedge }{B}_{\mu \nu }$ in order to build a nilpotent Slavnov operator \cite{livro}. In our case
we have an alternative allowed by the presence of the noncommutative $\theta 
$ parameter. Let us deform the transformation of the $\widehat{B}_{\mu \nu }$
field in $(\ref{setbrs})$ as follows

\begin{equation}
\stackrel{\wedge }{s}\stackrel{\wedge }{B}_{\mu \nu }=\stackrel{\wedge }{D}%
_{[\mu }\stackrel{\wedge }{\psi }_{\nu ]}+i\left[ \stackrel{\wedge }{c}\stackrel%
{\ast }{,}\stackrel{\wedge }{B}_{\mu \nu }\right] +\stackrel{\wedge }{\delta }%
_{t}\stackrel{\wedge }{B}_{\mu \nu }\mathrm{{\ },}  \label{simb}
\end{equation}%
in such a way that the nilpotency of $\stackrel{\wedge }{s}$ can be recovered

\begin{equation}
\stackrel{\wedge }{s}^{2}\stackrel{\wedge }{B}_{\mu \nu }=-i\left[ \stackrel{%
\wedge }{F}_{\mu \nu }\stackrel{\ast }{,}\stackrel{\wedge }{\rho }\right] 
\mathrm{\ }+\stackrel{\wedge }{s}(\stackrel{\wedge }{\delta }_{t}\stackrel{%
\wedge }{B}_{\mu \nu })-i\left[ \stackrel{\wedge }{c}\stackrel{\ast }{,}%
\stackrel{\wedge }{\delta }_{t}\stackrel{\wedge }{B}_{\mu \nu }\right] =0\ .
\label{recover}
\end{equation}%
At this point, we can see that we are generalizing the procedure that was
taken in the introduction of this work when we were dealing with the $\theta 
$ first order case (equation $(\ref{simb})$ is the all orders generalization
of $(\ref{newnctop})$). 

A brief comment is important here. In the BV language, in the presence of the anti-fields, the idea of deforming the set of field transformations of a given theory together with the deformation of its action is well known in the literature \cite{Henneaux}. In fact, equation $(\ref{recover})$ when translated to the BV language, would appear as the correction demanded by the Master Equation (it woud come from the term joining the transformation of the ghost $\stackrel{\wedge }{\psi }_{\mu }$ an of its anti-fields). This correction would just be the introduction of the quadratic term in the anti-field of $\stackrel{\wedge }{B}_{\mu \nu }$ in the BV action, as it happens in the commutative case that we have. But as we are seeking for a SW map transforming the fields of the noncommutative $\stackrel{\wedge }{BF}$ theory into those of the usual commutative theory, we avoid the introduction of the BV anti-fields. This is only possible due to the existence of the noncommutative parameter $\theta^{\mu\nu}$, which allows us to restrain ourselves to the set of the fields of the theory without the presence of the anti-fields. Then we find this present approach a more direct attack to the SW map problem.

The solution of (\ref{recover}) is thus the key ingredient to assert the
existence of the SW mapping, implying that 
\begin{equation}
\stackrel{\wedge }{s}(\stackrel{\wedge }{\delta }_{t}\stackrel{\wedge }{B}_{\mu
\nu })=i\left[ \stackrel{\wedge }{F}_{\mu \nu }\stackrel{\ast }{,}\stackrel{%
\wedge }{\rho }\right] +i\left[ \stackrel{\wedge }{c}\stackrel{\ast }{,}%
\stackrel{\wedge }{\delta }_{t}\stackrel{\wedge }{B}_{\mu \nu }\right] \mathrm{%
{\ }.}  \label{sdeltat}
\end{equation}%
We can now act with $\delta _{\theta }$ on this equation,%
\begin{eqnarray}
\stackrel{\wedge }{s}(\delta _{\theta }\stackrel{\wedge }{\delta }_{t}\stackrel{%
\wedge }{B}_{\mu \nu }) &=&i\left[ \delta _{\theta }\stackrel{\wedge }{F}%
_{\mu \nu }\stackrel{\ast }{,}\stackrel{\wedge }{\rho }\right] -\frac{1}{2}%
\delta \theta ^{\alpha \beta }\left\{ \partial _{\alpha }\stackrel{\wedge }{F}%
_{\mu \nu }\stackrel{\ast }{,}\partial _{\beta }\stackrel{\wedge }{\rho }%
\right\} +i\left[ \stackrel{\wedge }{F}_{\mu \nu }\stackrel{\ast }{,}\delta
_{\theta }\stackrel{\wedge }{\rho }\right]  \label{sdeltateta} \\
&&+i\left[ \delta _{\theta }\stackrel{\wedge }{c}\stackrel{\ast }{,}\stackrel{%
\wedge }{\delta }_{t}\stackrel{\wedge }{B}_{\mu \nu }\right] -\frac{1}{2}%
\delta \theta ^{\alpha \beta }\left\{ \partial _{\alpha }\stackrel{\wedge }{c}%
\stackrel{\ast }{,}\partial _{\beta }(\stackrel{\wedge }{\delta }_{t}\stackrel{%
\wedge }{B}_{\mu \nu })\right\} +i\left[ \stackrel{\wedge }{c}\stackrel{\ast }{%
,}\delta _{\theta }\stackrel{\wedge }{\delta }_{t}\stackrel{\wedge }{B}_{\mu
\nu }\right] \mathrm{{\ },}  \nonumber
\end{eqnarray}%
use $(\ref{deltac})$, $(\ref{deltaa})$, $(\ref{deltarho})$, and, then, find

\begin{eqnarray}
\delta _{\theta }\stackrel{\wedge }{\delta }_{t}\stackrel{\wedge }{B}_{\mu \nu
} &=&-\frac{1}{4}\delta \theta ^{\alpha \beta }\left\{ \stackrel{\wedge }{A}%
_{\alpha }\stackrel{\ast }{,}(\partial _{\beta }+\stackrel{\wedge }{D}_{\beta
})\stackrel{\wedge }{\delta }_{t}\stackrel{\wedge }{B}_{\mu \nu }\right\} -%
\frac{1}{2}\delta \theta ^{\alpha \beta }\left\{ \stackrel{\wedge }{D}%
_{\alpha }\stackrel{\wedge }{F}_{\mu \nu }\stackrel{\ast }{,}\stackrel{\wedge }{%
\psi }_{\beta }\right\}  \nonumber \\
&-&\frac{1}{2}\delta \theta ^{\alpha \beta }\{\stackrel{\wedge }{F}_{[\mu
\alpha }\stackrel{\ast }{,}\stackrel{\wedge }{\delta }_{t}\stackrel{\wedge }{B}%
_{\beta \nu ]}\}+\delta \theta ^{\alpha \beta }\stackrel{\wedge }{X}_{\mu \nu
\alpha \beta },  \label{sol}
\end{eqnarray}%
w\noindent here $\stackrel{\wedge }{X}_{\mu \nu \alpha \beta }$ represents
the freedom in the solution of this kind of problem by covariant terms, as
we saw in $(\ref{cov})$, 
\begin{equation}
\stackrel{\wedge }{s}\stackrel{\wedge }{X}_{\mu \nu \alpha \beta }=i\left[ 
\stackrel{\wedge }{c}\stackrel{\ast }{,}\stackrel{\wedge }{X}_{\mu \nu \alpha
\beta }\right] \mathrm{.}  \label{cov2}
\end{equation}%
But there are still some restrictions on the solution $(\ref{sol})$ . When
we considered the deformation of the transformation of $\stackrel{\wedge }{B}%
_{\mu \nu }$ in $(\ref{simb})$, we obviously intended that it would still
represent a symmetry of the action. This implies that (taking $\stackrel{%
\wedge }{\delta }_{t}\stackrel{\wedge }{A}_{\mu }=0$) 
\begin{equation}
\stackrel{\wedge }{\delta }_{t}\mathcal{S}_{\widehat{BF}}=\int d^{4}x\frac{1}{%
4}\varepsilon ^{\mu \nu \varrho \sigma }\stackrel{\wedge }{F}_{\mu \nu }%
\stackrel{\wedge }{\delta }_{t}\stackrel{\wedge }{B}_{\varrho \sigma }=0%
\mathrm{,}  \label{deltas}
\end{equation}%
and 
\begin{equation}
\delta _{\theta }\stackrel{\wedge }{\delta }_{t}\mathcal{S}_{\widehat{BF}%
}=\int d^{4}x\frac{1}{4}\varepsilon ^{\mu \nu \varrho \sigma }(\delta
_{\theta }\stackrel{\wedge }{F}_{\mu \nu }\stackrel{\wedge }{\delta }_{t}%
\stackrel{\wedge }{B}_{\varrho \sigma }+\stackrel{\wedge }{F}_{\mu \nu }\delta
_{\theta }\stackrel{\wedge }{\delta }_{t}\stackrel{\wedge }{B}_{\varrho \sigma
})=0\mathrm{.}  \label{double}
\end{equation}%
Substituting $(\ref{sol})$ in $(\ref{double})$ we get

\begin{equation}
\int d^{4}x\varepsilon ^{\mu \nu \rho \sigma }\delta \theta ^{\alpha \beta
}\left( \frac{1}{2}\left\{ \stackrel{\wedge }{F}_{\mu \nu }\stackrel{\ast }{,}%
\stackrel{\wedge }{F}_{\alpha \rho }\right\} \stackrel{\wedge }{\delta }_{t}%
\stackrel{\wedge }{B}_{\beta \sigma }-\frac{1}{2}\stackrel{\wedge }{F}_{\mu
\nu }\left\{ \stackrel{\wedge }{D}_{\alpha }\stackrel{\wedge }{F}_{\rho \sigma
}\stackrel{\ast }{,}\stackrel{\wedge }{\psi }_{\beta }\right\} +\stackrel{%
\wedge }{F}_{\rho \sigma }\stackrel{\wedge }{X}_{\mu \nu \alpha \beta
}\right) =0  \label{vari1}
\end{equation}%
which can be rewritten as 
\begin{equation}
\int d^{4}x\varepsilon ^{\mu \nu \rho \sigma }\delta \theta ^{\alpha \beta }%
\stackrel{\wedge }{F}_{\rho \sigma }\left( \stackrel{\wedge }{X}_{\mu \nu
\alpha \beta }+\frac{1}{2}\left\{ \stackrel{\wedge }{F}_{\alpha \mu }\stackrel{%
\ast }{,}\stackrel{\wedge }{\delta }_{t}\stackrel{\wedge }{B}_{\beta \nu
}\right\} -\frac{1}{2}\left\{ \stackrel{\wedge }{D}_{\alpha }\stackrel{\wedge }%
{F}_{\mu \nu }\stackrel{\ast }{,}\stackrel{\wedge }{\psi }_{\beta }\right\}
\right) =0\mathrm{{}.}  \label{vari2}
\end{equation}%
Before the final identification, we have to take care to guarantee the
covariance of $\stackrel{\wedge }{X}_{\mu \nu \alpha \beta }$. Remembering
the transformation of $\stackrel{\wedge }{\delta }_{t}\stackrel{\wedge }{B}%
_{\mu \nu }$ in $(\ref{sdeltat})$, it is not difficult to change the form of
the last term in $(\ref{vari2})$, integrating by parts and using a Fierz
identity, to arrive at 
\begin{equation}
\delta \theta ^{\alpha \beta }\stackrel{\wedge }{X}_{\mu \nu \alpha \beta }=%
\frac{1}{4}\delta \theta ^{\alpha \beta }\left\{ \stackrel{\wedge }{F}_{[\mu
\alpha }\stackrel{\ast }{,}\stackrel{\wedge }{\delta }_{t}\stackrel{\wedge }{B}%
_{\beta \nu ]}\right\} +\frac{1}{4}\delta \theta ^{\alpha \beta }\left\{ 
\stackrel{\wedge }{F}_{[\mu \alpha }\stackrel{\ast }{,}\stackrel{\wedge }{D}%
_{\beta }\stackrel{\wedge }{\psi }_{\nu ]}-\stackrel{\wedge }{D}_{\nu ]}%
\stackrel{\wedge }{\psi }_{\beta }\right\} \mathrm{{\ }.}  \label{varix}
\end{equation}%
(It must be emphasized that the solution of $(\ref{varix})$ is still a
particular solution of $(\ref{vari2})$. It is always possible to introduce
covariant terms satisfying 
\begin{equation}
\int \varepsilon ^{\mu \nu \rho \sigma }\delta \theta ^{\alpha \beta }%
\stackrel{\wedge }{F}_{\mu \nu }\stackrel{\wedge }{X^{\prime }}_{\rho \sigma
\alpha \beta }=0  \label{covaterm}
\end{equation}%
with a covariant $\stackrel{\wedge }{X^{\prime }}_{\rho \sigma \alpha \beta }$%
. We will not explore this freedom)

At this point, we can see that the extended symmetry $(\ref{newtop})$ that
we have found at the first explicit order in $\theta $, necessary for the
construction of the SW map, is just the first order integral of $(\ref{sol})$
with $(\ref{varix})$.

Finally, we can search for $\delta _{\theta }\stackrel{\wedge }{B}_{\mu \nu }$%
. Applying $\delta _{\theta }$ on $(\ref{simb})$,

\begin{eqnarray}
\delta _{\theta }\stackrel{\wedge }{s}\stackrel{\wedge }{B}_{\mu \nu } &=&%
\stackrel{\wedge }{D}_{[\mu }\delta _{\theta }\stackrel{\wedge }{\psi }_{\nu
]}-i\left[ \delta _{\theta }\stackrel{\wedge }{A}_{[\mu }\stackrel{\ast }{,}%
\stackrel{\wedge }{\psi }_{\nu ]}\right] +\frac{1}{2}\delta \theta ^{\alpha
\beta }\left\{ \partial _{\alpha }\stackrel{\wedge }{A}_{[\mu }\stackrel{\ast }%
{,}\partial _{\beta }\stackrel{\wedge }{\psi }_{\nu ]}\right\} +i\left[
\delta _{\theta }\stackrel{\wedge }{c}\stackrel{\ast }{,}\stackrel{\wedge }{B}%
_{\mu \nu }\right]  \nonumber \\
&-&\frac{1}{2}\delta \theta ^{\alpha \beta }\left\{ \partial _{\alpha }%
\stackrel{\wedge }{c}\stackrel{\ast }{,}\partial _{\beta }\stackrel{\wedge }{B}%
_{\mu \nu }\right\} +i\left[ \stackrel{\wedge }{c}\stackrel{\ast }{,}\delta
_{\theta }\stackrel{\wedge }{B}_{\mu \nu }\right] +\delta _{\theta }\stackrel{%
\wedge }{\delta }_{t}\stackrel{\wedge }{B}_{\mu \nu }\mathrm{{\ },}
\label{finalvari}
\end{eqnarray}%
and then using $(\ref{deltac})$, $(\ref{deltaa})$, $(\ref{deltapsi})$, $(\ref%
{sol})$, and $(\ref{varix})$, we can solve for $\delta _{\theta }\stackrel{%
\wedge }{B}_{\mu \nu }$,%
\begin{equation}
\delta _{\theta }\stackrel{\wedge }{B}_{\mu \nu }=-\frac{1}{4}\delta \theta
^{\alpha \beta }\left\{ \stackrel{\wedge }{A}_{\alpha }\stackrel{\ast }{,}%
(\partial _{\beta }+\stackrel{\wedge }{D}_{\beta })\stackrel{\wedge }{B}_{\mu
\nu }\right\} -\frac{1}{4}\delta \theta ^{\alpha \beta }\left\{ \stackrel{%
\wedge }{F}_{[\mu \alpha }\stackrel{\ast }{,}\stackrel{\wedge }{B}_{\beta \nu
]}\right\} \mathrm{{\ }.}  \label{deltab}
\end{equation}%
This is the relation that was missing to complete the dependence of the
fields of the $\widehat{BF}$ theory on $\theta $. We can also observe that $(%
\ref{b1order})$ is the first order solution of $(\ref{deltab})$, as expected.

\section{Conclusion}

Once we obtained the dependence on $\theta $ of the fields of the
noncommutative $\widehat{BF}$ theory after a SW map, we can answer the
question of what is the behavior of the action $(\ref{ncbf})$ under such
map. Its dependence on $\theta $ can be obtained by applying $\delta
_{\theta }$ on $(\ref{ncbf})$, and after using $(\ref{deltaa})$ and $(\ref%
{deltab})$, we find

\begin{equation}
\delta _{\theta }\mathcal{S}_{\widehat{BF}}=\frac{1}{8}\int
d^{4}x\varepsilon ^{\mu \nu \rho \sigma }\delta \theta ^{\alpha \beta
}\left( \left\{ \stackrel{\wedge }{F}_{\alpha \mu }\stackrel{\ast }{,}\stackrel{%
\wedge }{F}_{\beta \nu }\right\} \stackrel{\wedge }{B}_{\rho \sigma }-\frac{1%
}{2}\left\{ \stackrel{\wedge }{F}_{\alpha \beta }\stackrel{\ast }{,}\stackrel{%
\wedge }{F}_{\mu \nu }\right\} \stackrel{\wedge }{B}_{\rho \sigma }+\stackrel{%
\wedge }{F}_{\mu \nu }\left\{ \stackrel{\wedge }{F}_{\alpha \rho }\stackrel{%
\ast }{,}\stackrel{\wedge }{B}_{\beta \sigma }\right\} \right) \mathrm{{\ }.}
\label{deltancbf}
\end{equation}%
By means of a Fierz identity, we can show that the expression on the right
hand side of $(\ref{deltancbf})$ is null:%
\begin{equation}
\delta _{\theta }\mathcal{S}_{\widehat{BF}}=0\mathrm{.}  \label{deltancbf0}
\end{equation}

The final conclusion is that, with the particular Seiberg-Witten
transformation determined by $(\ref{deltaa})$ and $(\ref{deltab})$, the
noncommutative $\widehat{BF}$ action is mapped into its pure commutative
version without any nonrenormalizable  corrections in $\theta $. The ambiguities in the SW map can generate
covariant deformations in $\theta $ in the commutative space. But as such
deformations only intervene in the interaction sector of the theory, without
changing the propagation sector, the topological variables of the $BF$
theory (linking numbers) will remain independent of $\theta $ without
feeling the presence of these deformations. This is in complete analogy with
the case of the noncommutative Chern-Simmons model in $3D$ \cite{gs}. We
believe that this renormalizability is a general feature of the commutative theories obtained through the SW map of noncommutative actions with
Schwarz type topological sectors.

\section*{Acknowledgments}

The Conselho Nacional de Desenvolvimento Cient\'{\i}fico e Tecnol\'{o}gico
(CNPq-Brazil), the Funda{\c{c}}{\~{a}}o de Amparo \`a Pesquisa do Estado do
Rio de Janeiro (Faperj) and the SR2-UERJ are acknowledged for financial
support.

\end{document}